\documentclass{article}
\usepackage{latexsym}
\usepackage{amssymb}
\usepackage{amsmath}

\begin{document}

\title{Thomas rotation and Thomas precession}
\author{Tam\'as Matolcsi -- M\'at\'e Matolcsi \\  Department of Applied Analysis, 
E\"otv\"os Lor\'and University \\ Budapest}
\maketitle

\newcommand\uu{\mathbf u} \newcommand\uo{\mathbf u_0}
\newcommand\Eu{\mathbf E_\uu} \newcommand\Eui{\mathbf E_\ui}
\newcommand\x{\mathbf x}\newcommand\y{\mathbf y}
\newcommand\I{\mathbf I}\newcommand\h{\mathbf h}
\newcommand\M{\mathbf M} \newcommand\B{\mathbf B} 
\renewcommand\t{\mathbf t}\newcommand\ui{\mathbf u'}
\renewcommand\l{\lambda}\newcommand\piu{\boldsymbol\pi_\uu}
\newcommand\s{\mathbf s}\newcommand\F{\mathbf F}
\newcommand\be{\begin{equation}}
\newcommand\U{\mathbf U}
\newcommand\R{\mathbf R}
\newcommand\q{\mathbf q}
\newcommand\Om{\boldsymbol\Omega}
\newcommand\E{\mathbf E}
\newcommand\z{\mathbf z}
\newcommand\A{\mathbf A}
\newcommand\uc{\uu_c}\newcommand\e{\mathbf e}
\newcommand\rr{\mathbb{R}}
\newcommand\vv{\mathbf v}\newcommand\as{\mathbf a}

\newpage
\begin{abstract}
Exact and simple calculation of Thomas rotation and Thomas precessions 
along a circular world line is presented
in an absolute (coordinate-free) formulation of
special relativity. Besides the simplicity of calculations the absolute 
treatment of spacetime allows us to gain a deeper insight
into the phenomena of Thomas rotation and Thomas precession.
\end{abstract}


\baselineskip=20pt

\section{Introduction}

The 'paradoxic' phenomenon of Thomas precession has given rise to much 
discussion ever since the publication of Thomas' seminal paper 
(Thomas 1927) in which he made a correction by a factor 1/2 to the 
angular velocity of the spin of an electron moving in a magnetic field.  
Let us mention here that in the literature 
there seems to be no standard agreement as to the usage of the terms 
'Thomas precession' and 'Thomas rotation'. 
As explained in more detail in Section \ref{s:thprec} below, we prefer to 
use the term {\it Thomas precession} to refer to the continuous change of 
direction, {\it with respect to an inertial frame},
of a gyroscopic vector moving along a world line. {\it Thomas rotation}, 
on the other hand, will refer to the spatial rotation experienced by a 
gyroscopic vector having moved along a 'closed' 
world line, and having returned to its initial frame of reference (see Section
\ref{s:gen}).
       
One of the most studied cases (see e.g. Costella et al. 2001, Kennedy 2002) 
is the fact that the application of three successive
Lorentz boosts (with the relative 
velocities adding up to zero) results, in general, in a spatial rotation: the 
{\it discrete} Thomas rotation  
(see Section \ref{s:finite} for details). The same fact is often described as 
'the composition of two Lorentz boosts is 
equivalent to a boost and a spatial rotation'. We prefer to use three 
Lorentz boosts instead (with the relative 
velocities adding up to zero), in order to return to the initial frame of 
reference, in accordance with our terminology of Thomas 
rotation. Describing the mathematical structure of discrete 
Thomas rotations has motivated A.A. Ungar to build the comprehensive 
theory of gyrogroups and gyrovector spaces (Ungar 2001). 

The other case typically under consideration comes from the original 
observation of Thomas: the continuous change of direction, with 
respect to an inertial frame, of a  gyroscopic vector moving along a 
circular orbit. This phenomenon 
has been subject to considerations from various points of view 
(Muller 1992 (Appendix), 
Philpott 1996, Rebilas 2002 (Appendix), Herrera \& Di Prisco 2002, 
Rhodes \& Shemon 2003). 
The considerations usually involve, either explicitly or implicitly, the viewpoint of the orbiting 'airplane', i.e. 
a rotating observer. This might lead us to believe  (see Herrera \& Di Prisco 2002) 
that the calculated angle of rotation depends on the definition of the rotating observer (and this could lead to 
an experimental checking of what the 'right' definition of a rotating observer is). From our treatment below, however, 
it will be clear the Thomas rotation is an absolute fact,  independent of the rotating (or, any other) observer.  

It is also interesting to note that new connections 
between quantum mechanical phenomena and Thomas rotation have recently been 
pointed out (L\'evay 2004).

As it is well known, the theory of special relativity contradicts our 
common sense notions about space and time in many respects. Early day 
'paradoxes' were usually based on our intuitive
assumption of absolute simultaneity. With the resolution of paradoxes 
such as the 'twin paradox' or the 'tunnel paradox' it has become common 
knowledge that the concept of time must be handled very carefully. 
As it is also well known, 
the theory of special relativity implies, besides the non-existence of absolute time,  the {\it non-existence of 
absolute space}. An expression such as 'a point in space' simply does not have an 
absolute meaning, just as the expression 'an instant in time'.  However, this fact seems to be given less attention to 
and even overlooked sometimes. The fact that the space vectors of any observer 
are usually represented as vectors in $\rr^3$ leads one to forget that 
these spaces really are {\it different}. This conceptual error lead e.g. 
to the 'velocity addition paradox' (Mocanu 1992). The spaces of two 
different inertial observers are, of course, connected via the corresponding Lorentz 
boost, and the non-transitivity of Lorentz boosts (which, in fact,
gives rise to the notion of Thomas rotation) gave the correct explanation 
of this 'paradox' (Ungar 1989, Matolcsi \& Goher 2001).

To grasp the essence of the concepts related to Thomas rotation, let us mention that 
in some sense this intriguing phenomenon is analogous to the 
well known twin paradox. Consider two twins 
in an inertial frame. One of them remains in that frame for all times, 
while the other goes for a trip in spacetime, and later returns to his 
brother. It is well-known that different times have passed for the two twins: 
the traveller is younger than his brother. 
What may be surprising is that the space of the traveller when 
he arrives, although he experienced no torque during his journey,
will be rotated compared to the space of his brother; this is, 
in fact, the Thomas rotation. This analogy is illuminating in one 
more respect: until the traveller returns to the original frame of 
reference it makes no sense to ask  'how much younger is the traveller compared to 
his brother?' and 'by what angle is the traveller's gyroscope 
rotated compared to that of his brother?' Different observers may give different answers.
When the traveller returns to his brother, 
these questions suddenly make perfect sense, and there is an absolute answer (independent of who the observer is)
as to how much younger and how much rotated the traveller is.

Of course, an arbitrtary inertial frame can observe the brothers continuously, and can tell,
at each of the frame's instants, what difference {\it he sees} between the
ages of the brothers. More explicitly, as it is well known, given a world line,
an arbitrary inertial frame can tell the relation between the frame's time and 
the proper time of the world line. This relation depends on the 
inertial frame: different inertial frames establish different relations.

Similarly, an arbitrary inertial frame, observing the two brothers, can tell
at each frame-instant what  difference {\it he sees} between the 
directions of the gyroscopes of the brothers. Different inertial frames
establish different relations.

This philsophy makes a clear distinction between 
Thomas rotation and Thomas precession connected to a world line: 

-- \ Thomas rotation refers to an absolute fact (independent of who observes it), which makes sense only 
for two equal local rest frames (if such exist) of the world line,

-- \ Thomas precession refers to a relative fact (i.e. depending on who observes the motion), which makes sense 
with respect to an arbitrary inertial frame.
   
In this paper we use the formalism of (Matolcsi 1993) to give a concise 
and rigorous treatment of the discrete and circular-path Thomas rotations. 
The Thomas rotation as well as the Thomas precession (with respect 
to certain inertial observers) along a circular world line are calculated. 
Our basic concept here is that special relativistic spacetime has a 
four-dimensional affine structure, and coordinatization 
(relative to some observer) is, in many cases, unnecesary in the description 
of physical phenomena. In fact, coordinates can sometimes lead to ambiguities 
in concepts and definitions, and bear the danger of leading us to overlook 
the fact that absolute space does not exist. 

As well as providing a clear overview of the appearing concepts, the 
coordinate-free formulation of special 
relativity enables us to give simple calculations. The indispensable 
Fermi-Walker equation is also straightforward to derive in our formalism.

\section{Fundamental notions}

In this section some notions and results of the special
relativistic spacetime model as a mathematical 
structure (Matolcsi 1993, 1998, 2001) will be recapitulated.
As the formalism slightly differs from the usual 
textbook treatments of special relativity 
(but only the formalism: our treatment is mathematically equivalent to 
the usual treatments), we will point out several relations 
between textbook formulae and those of our formalism. 

Special relativistic spacetime is an oriented four dimensional
affine space $M$ over the vector space $\M$; the spacetime
distances form an oriented one dimensional vector space $\I$,
and an arrow oriented Lorentz form $\M\times\M\to\I\otimes\I$,
$(\x,\y)\mapsto\x\cdot\y$ is given. 

An {\it absolute velocity} $\uu$ is a future directed
element of $\frac{\M}{\I}$ for which $\uu\cdot\uu=-1$ holds
(absolute velocity corresponds to four-velocity in usual
terminology).

For an absolute velocity $\uu$, we define the three 
dimensional spacelike linear subspace 
\be
\Eu:=\{\x\in\M\mid \uu\cdot\x=0\};
\end{equation}
then 
\be 
\piu:= 1+\uu\otimes\uu: \M\to \Eu,\quad \x\mapsto\x+\uu(\uu\cdot\x)
\end{equation}
is the projection onto $\Eu$ along $\uu$. The restriction of 
the Lorentz form onto $\Eu$ is positive definite, so $\Eu$
is a Euclidean vector space (this will correspond to the space vectors 
of an inertial observer with velocity $\uu$).

The history of a classical material point is described 
by a differentiable {\it world line function} $r:\I\to M$
such that $\dot r(\s)$ is an absolute velocity for all proper 
time values $\s$. The range of a world line function -- 
a one dimensional submanifold -- is called a {\it world line}.

An {\it observer} $\U$ is an absolute velocity valued smooth map
defined in a connected open subset of $M$. (This is just a mathematical
 definition; it may sound unfamiliar at first, but considering that 
something that an observer calls a 'fixed space-point' 
is, in fact, a world line in spacetime,
this definition will make perfect 'physical' sense).
A maximal integral curve of $\U$ -- a world line -- is a 
{\it space point} of the observer, briefly a $\U$-{\it space point}; 
the set of the maximal integral curves of $\U$ is the {\it space} of 
the observer, briefly the $\U$-{\it space}.  

An observer having constant value is called
{\it inertial}. An inertial observer will be referred to by
its constant velocity.
The space points -- the integral curves -- of
an inertial observer with absolute velocity $\uu$ are straight 
lines parallel to $\uu$. The $\uu$-space point containing the
world point $x$ is the straight line $x+\uu\I$, where
$\uu\I:=\{\uu\t\mid \t\in\I\}$.

In order to arrive at the analogue of the coordinate system 
corresponding to an inertial observer 
we need to specify the time-syncronization of the observer. Of course, the 
standard syncroniztion is used: 
according to the standard synchronization of $\uu$, two world
points  $x$ and $y$ are simultaneous if and only if
$\uu\cdot(y-x)=0$. Thus, simultaneous world points form a
hyperplane parallel to $\Eu$; such a hyperplane is
an $\uu$-{\it instant}, their set is $\uu$-{\it time}. 
The $\uu$-instant containing the world point $x$ is the
hyperplane $x+\Eu$.

An inertial observer together with its standard
synchronization is called a {\it standard inertial frame}. 
Note that a standard inertial frame is an exactly defined
object in our framework, it does not refer to any coordinates,
coordinate axes, it contains an inertial observer and its 
standard synchronization only. 

The space vector between two $\uu$-space points (straight lines
in spacetime) is the world vector between 
$\uu$-simultaneous world points of the straight lines in question;
in formula, $\uu$-space, endowed with the subtraction
\be 
(x+\uu\I)-(y+\uu\I):=\piu(x -y)\label{terkiv}
\end{equation} 
becomes a three dimensional affine space over $\Eu$ (this fact shows 
that $\Eu$ does indeed 
correspond to the space vectors of the observer $\uu$). 

This is a crucial point: the space vectors of the standard
inertial frame $\uu$ are elements of $\Eu$, so {\it the space 
vectors of different inertial frames form different three dimensional 
vector spaces.}

The time passed between to $\uu$-instants (hyperplanes in
spacetime) is the time passed between them in an arbitrary 
$\uu$-space point. In formula, $\uu$-time, endowed with the subtraction
\be 
(x+\Eu)-(y+\Eu):=-\uu\cdot(x-y) \label{idokiv} 
\end{equation} 
becomes a one dimensional affine space over $\I$.

If $r$ is a world line function, then the standard inertial frame
with velocity value $\dot r(\s)$ is called the {\it local
rest frame corresponding to $r$ at $\s$}.

In usual treatments the coordinates distinguish a certain inertial 
frame (the 'rest' frame) and any other inertial frame is considered 
through its relative velocity with respect to the rest frame (and the 
coordinates with respect to the new frame are given via the 
corresponding Lorentz transformation).
The main feature of our approach is the systematic use of
absolute velocities for characterizing standard inertial frames 
(this perfectly reflects the principle of relativity: no inertial frame can 
be distinguished compared to other inertial frames).
Among several advantages, such as clarity of many concepts appearing in 
the theory of relativity, it often results in 
highly simplified and clear formulae. 

\section{Relative velocity and relative acceleration}\label{s:relva}

Let $r$ be a world line function $r$ (describing the history of a classical
material point). A standard inertial frame with absolute velocity $\uu$
 gives a correspondence between 
$\uu$-time $t$ and the proper time $\s$ of the world line function $r$:
if $t_0$ is the  $\uu$-instant of the world point $r(0)$, then, according
to (\ref{idokiv}), $\t:=(r(\s)+\Eu)-(r(0)+\Eu)=-\uu\cdot(r(\s)-r(0))$; 
therefore
\be\label{idok}
\frac{d\t}{d\s} =-\uu\cdot\dot r(\s).
\end{equation}
As a consequence, the proper time, too, can be given as a function
of $\uu$-time, and
\be
\frac{d\s}{d\t} =\frac1{-\uu\cdot\dot r(\s(\t))}.
\end{equation}

The inertial frame observes the history of the material point as a motion, 
assigning $\uu$-space points to $\uu$-instants : $r_\uu(\t):= r(\s(\t))+
\uu\I$.
Then, according to (\ref{terkiv}) and the previous equality,
the relative velocity is (for the sake of brevity we omit the variable $\t$ from the 
expressions)
\begin{equation}\label{relvel}
\vv_\uu:= r_\uu'= \lim_{\h\to0}\frac{r_\uu(\t+\h)-r_\uu(\t)}
{\h}=
\frac{\dot r(\s)}{-\uu\cdot\dot r(\s)} -\uu
\end{equation}
and the relative acceleration is
\begin{equation}\label{relac}
\as_\uu:= r_\uu''= \frac1{(-\uu\cdot\dot r(\s))^2}\left(\ddot r(\s)
+\frac{\dot r(\s)(\uu\cdot\ddot r(\s)}{-\uu\cdot\dot r(\s)}\right)
\end{equation}
where the derivative according to $\uu$-time is denoted by a prime.

It is worth mentioning that
\be\label{relfactor}
-\uu\cdot \dot r(\s)=\frac1{\sqrt{1-|\vv_\uu|^2}}=:\gamma_\uu,
\end{equation}
the well-known relativistic factor.

\section{Lorentz boosts and discrete Thomas rotations}\label{s:finite}

As we emphasized, the space vectors of different standard 
inertial frames form different three dimensional vector spaces;
for the absolute velocities $\uu$ and $\ui$, $\Eu$ and
$\Eui$ are different vector spaces. A natural correspondence
can be given between them, the {\it Lorentz boost} from
$\uu$ to $\ui$ (Matolcsi 1993, 2001),
\be
\B(\ui,\uu):= 1 + \frac{(\ui+\uu)\otimes(\ui+\uu)}{1-\ui\cdot\uu}
-2\ui\otimes\uu 
\end{equation}
which is a Lorentz form preserving linear map on $\M$, such that $B(\ui , \uu)\uu =\ui$.
This is the absolute form (which appears implicitly in Rowe 1984,
too) of the usual Lorentz boost. It is clear from the given formula that 
this absolute form depends on two absolute 
velocities. The explicit matrix form of a 
textbook Lorentz boost depends on a single relative velocity
but, in fact, it also refers to two inertial 
observers (one of which is the 'rest frame',  not appearing explicitly in the formulae).  
  
The vector $\q'$ in the space of the inertial frame $\ui$ is called
{\it physically equal} to the vector $\q$ in the space of the
inertial frame $\uu$ if $\q'=\B(\ui,\uu)\q$; we say also that
$\q$ {\it boosted} from $\Eu$ to $\Eui$ equals $\q'$. 
This Lorentz boost gives sense to the usual tacit assumption
that the corresponding coordinate axes of different inertial frames are
parallel. The coordinate axes defined by the vectors $\e_i$
in $\Eu$ are parallel to the axes defined by the vectors $\e'_i$
in $\Eui$ if $\e'_i=\B(\ui,\uu)\e_i$ \ $(i=1,2,3)$.
(The parallelism of frame axes is usually a nagging problem in standard treatments; see the 
discussion in the Introduction of Kennedy 2002.)

To be physically equal is a symmetric relation:  $\B(\ui,\uu)^{-1}=
\B(\uu,\ui)$, so if $\q'$ is physically equal to $\q$, then
$\q$ is physically equal to $\q'$.

On the other hand, to be physically equal is {\it not transitive}: 
the product of two Lorentz boosts, 
in general, is not a Lorentz boost (as it is well known): we have
\be
\B(\uu'',\ui)\B(\ui,\uu)=\B(\uu'',\uu)\quad \text{iff}
\quad \uu,\ui,\uu''\quad \text{are coplanar},
\end{equation}
(which is equivalent to the standard formalism:
the relative velocity of $\uu''$ with respect to $\uu$ and
the relative velocity of $\ui$ with respect to $\uu$ are collinear.)

In an equivalent formulation,
\be
\R_\uu(\ui,\uu''):=\B(\uu,\uu'')\B(\uu'',\ui)\B(\ui,\uu)
\end{equation}
is the identity transformation if and only if
$\uu,\ui,\uu''$ are coplanar.  
Note that $\R_\uu(\ui,\uu'')\uu=\uu$ and
the restriction of $\R_\uu(\ui,\uu'')$ onto $\Eu$ is a rotation,
called the {\it discrete Thomas rotation  corresponding to  $\uu$, $\ui$
and $\uu''$}. 

Thus if $\q'$ is physically equal to $\q$ and $\q''$ is physically equal
to $\q'$, then $\q$ need not be physicllay equal to $\q''$. 
This is why the Thomas rotation appears to be 'paradoxic'.

In other words, a vector $\q$ boosted from $\Eu$ to $\Eui$ yields $\q'$ and
then $\q'$ boosted from $\Eui$ to $\E_{\uu''}$ yields $\q''$, and
lastly $\q''$ boosted from $\E_{\uu''}$ back to $\Eu$, results in a 
vector rotated from the original $\q$. 

\section{Compasses}

A boost, as defined above, does not mean a real transport of vectors
from an observer space into another one. Nevertheless, it can
be related to such a transport in the following situation. 

A {\bf compass} (a needle fixed to a central point) can be
described in spacetime as a vector attached to a material point; 
more precisely, as a pair of functions $(r,\z)$ where
$r$ is a world line function (the history of the material point) and 
$\z$ is a vector valued function (describing the direction of the needle) 
defined  on the proper time of $r$, $\z:\I\to \M$, such that

-- \ it is always spacelike according to the corresponding local rest frame
of the world line, i.e. $\dot r\cdot\z=0$,

-- \ the magnitude of $\z$, $|\z|$ is constant.

Thus the needle of the compass passes continuously from the space
of one local rest frame to that of another one. The compass is conceived
to be {\bf locally inertial} if $\z$ is physically constant
along $r$ (keeps direction in itself) i.e. the values of $\z$ 
are boosted continuously 
corresponding to the absolute velocities of the world line. 
This means that if $\h$ is a "small" time period,
then  $\z(\s+\h)$ in $\E_{\dot r(\s+\h)}$ 
is "nearly" physically equal to $\z(\s)$ in $\E_{\dot r(\s)}$, 
more precisely
\be
\lim_{\h\to0}\frac{\z(\s+\h)-\B(\dot r(\s+\h),\dot r(\s))\z(\s)}
{\h}=0.
\end{equation}

Because $\dot r\cdot\z=0$, we can replace $\bigl(\dot r(\s+\h) + 
\dot r(\s)\bigr)\cdot\z(\s)$ with $\bigl(\dot r(\s+\h) - 
\dot r(\s)\bigr)\cdot\z(\s)$, so
\be
\B(\dot r(\s+\h),\dot r(\s))\z(\s)= \z(\s) + \frac{\bigl(\dot r(\s+\h) +
\dot r(\s)\bigr)\bigl(\dot r(\s+\h) - \dot r(\s)\bigr)\cdot\z(\s)}
{1 - \dot r(\s+\h)\cdot \dot r(\s)}
\end{equation}
and the above limit becomes $\dot\z - \dot r(\ddot r\cdot \z)=0$, from
which, taking into account again $\dot r\cdot\z=0$, we get the
well known Fermi-Walker equation along $r$
\be
\dot\z = \dot r(\ddot r\cdot \z) - \ddot r(\dot r\cdot\z) =
(\dot r\land\ddot r)\z.\label{gyreq}
\end{equation}

Note that the Lorentz boosts in terms of absolute velocities yielded this
equation in an extremely brief and simple way (in contrast to
the usual deductions, see e.g. M\o ller 1972).

If $\z$ is any vector satisfying the Fermi-Walker equation along $r$, 
then $(\dot r\cdot\z)\dot{}=0$,
so $\dot r\cdot\z$ is constant; if  $\z(\s_0)$ is spacelike according to 
$\dot r(\s_0)$ for one proper time value $\s_0$, then $\z(\s)$ is 
spacelike according to $\dot r(\s)$ for all $\s$ ($\z$ is always
spacelike according to the corresponding local rest frame of $r$).
Moreover, then  $\dot\z\cdot\z=0$, so the magnitude of $\z$
is constant.

Let us introduce another term. Let $r$ be world line function. We 
call a function $\z:\I\to\M$ a {\bf gyroscopic vector on} $r$
if $\z$ satisfies the Fermi-Walker equation along $r$ and a
value of $\z$ is spacelike according to the corresponding
local rest frame of $r$. Obviously, if $\z$ is a gyroscopic
vector along $r$, then $(r,\z)$ is a locally inertial compass.
It is well known and easily verifiable that if $\z_1$ and $\z_2$
are gyroscopic vectors on the same world line, then $\z_1\cdot\z_2$
is constant (which corresponds to the fact that 'non-rotating' vectors retain their
relative angle).

\section{Circular world line}\label{s:circw}

Take a standard inertial frame with velocity value $\uc$. A circular motion
with respect to this frame can be given by 

-- \ its centre $q_c$ in $\uc$-space,

-- \ its angular velocity, an antisymmetric linear map $0\neq\Om:
\E_{\uc}\to\frac{\E_{\uc}}{\I}$ (usually one considers angular 
velocity as a spatial axial vector which, in fact, corresponds
to an antisymmetric tensor),

-- \ its initial position with respect to the centre, a vector $0\neq\q$ 
in $\E_{\uc}$, orthogonal to the kernel of $\Om$  such that $|\Om\q|<1.$

This motion has the form 
\be
\t\mapsto q_c+e^{\t\Om}\q=q_c+\q\cos\omega\t +
\frac{\Om\q}{\omega}\sin\omega\t
\end{equation}

where $\omega:=|\Om|=\sqrt{\frac{1}{2} \mathrm{Tr}\Om^\ast \Om}$. Note that we have 
\be
\Om^2\q=-\omega^2\q,\qquad |\Om\q|=\omega\rho
\end{equation} 
where $\rho:=|\q|$.

The relative velocity of this motion equals $e^{\t\Om}\Om\q$ 
which has the magnitude $\omega\rho$. Thus, we infer from \eqref{idok} and \eqref{relfactor},
 that
the relation between the proper time $\s$ of the world line and the $\uc$-time $\t$ 
is $\t=\s\l$, where 
\be
\l:=\frac1{\sqrt{1-\omega^2\rho^2}}.
\end{equation}
 Then we easily derive that this 
motion comes from the world line function
\be 
\s\mapsto r(\s) = o + \s\l\uc
+e^{\s\l\Om}\q\label{circline}
\end{equation}
where $o$ is a world point of the centre $q_c$ (which is a straight
line in spacetime). Then
\be \label{circva}
\dot r(\s) = \l(\uc +e^{\s\l\Om}\Om\q),\qquad
\ddot r(\s)
=-\l^2\omega^2e^{\s\l\Om}\q.
\end{equation} 

Note that $\uc$ is the absolute velocity of the centre and
$\uo:=\l(\uc+\Om\q)$ is the "initial" absolute velocity of
the world line.

\section{Gyroscopic vectors on a circular world line}\label{s:gyrcirc}

Introducing the variable $\t:=\l\s$  ($\uc$-time) and the function
$\hat\z(\t):=\z(\t/\l)$, then
omitting the "hat" for brevity, we get the Fermi-Walker 
diferential equation (\ref{gyreq}) along the above
circular world line in the form
\be  
\z'(\t) = -\l^2\omega^2 \bigl((\uc+
e^{\t\Om}\Om\q)\land(e^{\t\Om}\q)\bigr)\z(\t).
\end{equation}

In the sequel we find it convenient to consider $\Om$ as defined on the whole of 
$\M$ in such a way that $\Om\uc=0$. Then $\Om$ will be a Lorentz antisymmetric
linear map on the whole of $\M$, thus 
$e^{\t\Om}$ will preserve the Lorentz form (it will be a Lorentz 
transformation) for which  $e^{\t\Om}\uc=\uc$ holds.

Then we infer that $\mathbf a(\t):=
e^{-\t\Om}\z(\t)$ satisfies the
autonomouos linear differential equation
\be 
{\mathbf a}'(\t) =-\A\mathbf a(\t)
\end{equation}
where
\be \label{A}
\A:=\Om + \l^2\omega^2(\uc +\Om\q)\land\q = \l^2\Om +
\l^2\omega^2\uc\land\q
\end{equation}
where the latter equality relies on the simple fact that 
\begin{equation}\label{omq}
(\Om\q)\land\q=\rho^2\Om.
\end{equation}

As a consequence -- since $\mathbf a(0)=\z(0)$ --,
we get the solution of the Fermi-Walker differential equation
in the form
\be 
\z(\t)=e^{\t\Om}e^{-\t\A}\z(0).
\end{equation} 

Let us investigate the properties of 
\be
\F(\t):=e^{\t\Om}e^{-\t\A}
\end{equation}
which we call the {\it Fermi-Walker operator} at $\t=\l\s$,
$\s$ being a proper time point of the circular world line function.

Since $\A$ is an antisymmetric linear map, $e^{-\t\A}$ is a Lorentz 
transformation.
It is trivial that $\A\uo=0$, thus the restriction of $e^{-\t\A}$ 
onto the three dimensional Euclidean space $\E_{\uo}$ is a rotation.

We know that the restriction of the Lorentz transformation $e^{\t\Om}$ 
onto the Euclidean vector space $\E_{\uc}$ is a rotation. 

Thus $e^{\t\Om}e^{-\t\A}$, as a product of two Lorentz transformations,
is a Lorentz transformation, too. Its restriction onto $\E_{\uo}$ 
is a Euclidean structure preserving linear bijection from $\E_{\uo}$ onto 
$\E_{\dot r(\t)}$. This can be conceived as a spatial rotation {\it only} if $\dot r(\t)=\uo$
(otherwise it acts between different 
Euclidean spaces).  

\section{Thomas rotation on the circular world line}\label{s:thangle}

The absolute velocity of the circular world line is periodic, $\dot
r(\frac{2\pi}{\omega})=\dot r(0)=\uo$. Since 
$e^{\frac{2\pi}{\omega}\Om}$ is the identity map, we have for the 
corrresponding Fermi-Walker operator 
\be
\F\left(\frac{2\pi}{\omega}\right)= e^{-\frac{2\pi}{\omega}\A}
\end{equation}
whose restriction onto the Euclidean vector space $\E_{\uo}$ is a
rotation, called the {\it Thomas rotation on the 
circular world line} (\ref{circline}).

The angle of the Thomas rotation is $2\pi-\frac{2\pi}{\omega}|\A|$ 
where $|\A|$ is the magnitude of $\A$; $|\A|:= \sqrt{|\A\e_1|^2 + 
|\A\e_2|^2}$ where $\e_1$ and $\e_2$ are arbitrary $\uo$-spacelike unit vectors 
orthogonal to the kernel of $\A$ such that $\e_1\cdot\e_2=0$.

It is trivial from \eqref{A} that if $\mathbf a\in\E_{\uc}$ is in the kernel of $\Om$
-- i.e. $\Om \mathbf a=0$ and $\q\cdot\mathbf a=0$ --, then $\mathbf a$ is in
the kernel of $\A$, too.
Therefore, the intersection $E_{\uo}\cap E_{\uc}\cap \mathrm{Ker} \A \cap \mathrm{Ker} \Om$
is 1-dimensional. 

This means that we can choose $\e_1:=\frac{\q}{|\q|}$ and $\e_2:=
\l\left(\omega\rho\uc +\frac{\Om\q}{\omega\rho}\right)$ (it is easy to verify that all 
conditions imposed on $\e_1$ and $\e_2$ are satisfied). Thus,
$\A\e_1=\l\omega\e_2$, $\A\e_2=-\l\omega\e_1$,
which implies that $|\A|= \l\omega$.

As a consequence, the Thomas angle on the circular world line
equals
\be
2\pi\left(1-\frac1{\sqrt{1-\omega^2\rho^2}}\right)
\end{equation}
which is the well known result (Thomas 1927).

It is worth noting that the value of a gyroscopic vector 
after a whole revolution equals the original one if and only
if the gyroscopic vector is parallel to the kernel of $\Om$
i.e. is orthogonal to the plane of rotation in the 
space of the centre.

\section{Generalizations}\label{s:gen}

Besides deriving the Thomas angle on the circular world line in 
a short and transparent way, our method gives the Thomas
rotation itself and allows us a deeper insight into the nature
of gyroscopic vectors in general.

Let $r$ be an arbitrary world line function.  The solutions of 
the corresponding Fermi-Walker equation with various initial values 
give us a Fermi-Walker operator $\F(\s_2,\s_1)$, a Lorentz
transformation for all proper time points $\s_1$ and $\s_2$
such that
\be
\dot r(\s_2)=\F(\s_2,\s_1)\dot r(\s_1)
\end{equation}
and
\be
\z(\s_2)=\F(\s_2,\s_1)\z(\s_1)
\end{equation}
for an arbitrary gyroscopic vector $\z$ on $r$. 

Thus the
restriction of $\F(\s_2,\s_1)$ onto $\E_{\dot r(\s_1)}$ --
the space vectors of the local rest frame at $\s_1$ --
is a Euclidean structure preserving linear bijection 
onto $\E_{\dot r(\s_2)}$ -- the space
vectors of the local rest frame at $\s_2$.

In particular, if $\dot r(\s_2)=\dot r(\s_1)$, the 
restriction of $\F(\s_2,\s_1)$ onto $\E_{\dot r(\s_1)}$ is a rotation,
which we call the {\it Thomas rotation on the world
line $r$, corresponding to the proper time points 
$\s_1$ and $\s_2$}.

It is worth noting: {\it a Thomas rotation on a world line for 
two proper time values has a meaning only if the corresponding 
absolute velocites are equal}. Thus no Thomas rotation can be defined
on a world line if all its absolute velocity values are different.

\section{Thomas precession with respect to an inertial frame}
\label{s:thprec}

Now, let $\z$ be a gyroscopic vector on the world line function $r$.
An inertial frame $\uu$ observes $\z$ by boosting it continuously
to its own space, i.e. giving the function $\z_\uu:\I\to\Eu$
such that
\be
\z_\uu(\t):=\B(\uu,\dot r(\s(\t)))\z(\s(\t)).
\end{equation}
Then, omitting $\t$ as previously, we infer that
\begin{equation}
\begin{split}
\z_\uu'=\frac1{-\uu\cdot\dot r(\s)}\Bigl(\Bigl(&\frac{d}{d\s}
\B(\uu,\dot r(\s))\Bigr)\z(\s) + \B(\uu,\dot r(\s))\dot\z(\s)\Bigr)\\
=\frac1{-\uu\cdot\dot r(\s)}\Bigl(\Bigl(&\frac{d}{d\s}
\B(\uu,\dot r(\s))\Bigr)\B(\dot r(\s),\uu)\z_\uu + \\
&\B(\uu,\dot r(\s))(\dot r(\s)\land\ddot r(\s))\B(\dot r(\s),\uu)\z_\uu)
\Bigr)
\end{split}
\end{equation}

Omitting $\s$ for the sake of brevity, we get immediately that the 
second term above equals 
\be
\uu\land\left(\ddot r + \frac{\dot r(\uu\cdot\ddot r)}{1-\uu
\cdot\dot r}\right).
\end{equation}
As concerns the first term, a straightforward calculation yields that 
it equals
\be
\frac{\dot r\land\ddot r}{1-\uu\cdot\dot r} -\uu\land\ddot r -
2\uu\land\left(\frac{\dot r(\uu\cdot\ddot r)}{1-\uu \cdot\dot r}
\right)
\end{equation}
Taking into account (\ref{relvel}) and (\ref{relac}), finally we obtain
the known result
\be
\z_\uu'=\frac{\gamma_\uu^2}{1+\gamma_\uu}(\vv_\uu\land\as_\uu)\z_\uu.
\end{equation}

Thus the inertial frame $\uu$ sees the gyroscopic vector $\z$
-- which keeps direction in itself -- precessing, the angular velocity
of precession is the antisymmetric linear map (depending on $\uu$-time)
\be
\Omega_\uu:=\frac{\gamma_\uu^2}{1+\gamma_\uu}\vv_\uu\land\as_\uu=
\frac{\gamma_\uu-1}{|\vv_\uu|^2}\vv_\uu\land\as_\uu
:\Eu\to\frac{\Eu}{\I}.
\end{equation}

Call attention to the fact: {\it the same gyroscopic vector precesses 
differently to different inertial frames}. 

\section{Thomas precessions corresponding to a circular world line}

Let us consider the circular world line described in Section \ref{s:circw}).

Let us take the standard inertial frame of the centre i.e. the one 
with absolute velocity $\uc$. Then equalities in (\ref{circva}), (\ref{relvel})
and (\ref{relac}) yield
\be
\vv_{\uc}(\t)=e^{\t\Om}\Om\q,\qquad \as_{\uc}(\t)=-\omega^2e^{\t\Om}\q.
\end{equation}
Then $\vv_{\uc}(\t)\land\as_{\uc}(\t)=-\omega^2e^{\t\Om}((\Om\q)\land\q)
e^{-\t\Om}=-\omega^2\rho^2\Om$ because of (\ref{omq}).
Since $\omega^2\rho^2=|\vv_{\uc}|^2$, the angular velocity of the 
Thomas precession with respect to the "central frame" $\uc$ is constant 
in $\uc$-time, equalling
\be
\left(1-\frac1{\sqrt{1-\omega^2\rho^2}}\right)\Om.
\end{equation}  

Usual treatments consider exclusively this precession (M\o ller,
......) in connection with the circular world line i.e. the
Thomas precession with respect to the central frame. Of course,
there are other possibilities, too.

For instance, let us take the standard inertial frame in which the 
gyroscopic vector is at rest initially i.e. the one with absolute velocity
$\uo=\l(\uc +\Om\q)$.

Then 
\be
-\uo\cdot\dot r(\s)=\l^2(1-\omega^2\rho^2\cos\omega\l\s).
\end{equation}
Consequently, now the $\uo$-time $\t$ and the proper time
$\s$ have the relation $\t=\l^2\s-\l\omega^2\rho^2\sin\omega\l\s$.
Then in view of (\ref{relvel}), we find
\be
\vv_{\uo}=\l\frac{(\uc+e^{\l\s\Om}\Om\q)(1-\omega^2\rho^2)}
{1-\omega^2\rho^2\cos\omega\l\s}- \l(\uc+\Om\q)
\end{equation}
and a similar, more complicated formula gives $\as_{\uo}$, too;
as a consequence, the angular velocity of the Thomas precession
with respect to the inertial frame $\uo$ depends rather
intricately on $\uo$-time. For instance, if $n$ is an arbitrary 
natural number, then 

-- \ for $\uo$-instants given by $\l\s=\frac{2n\pi}{\omega}$,
the value of the relative velocity is zero, so the angular velocity 
of Thomas precession has zero value, too; 

-- \ for $\uo$-instants given by $\l\s=\frac{(2n-1)\pi}{\omega}$,
the relative velocity equals \newline
 $-\frac{2\l}{1+\omega^2\rho^2}
(\omega^2\rho^2\uc+\Om\q)$ and the relative acceleration is 
$\frac{(1-\omega^2\rho^2)\omega^2}{(1+\omega^2\rho^2)^2}\q$, so 
the angular velocity of Thomas precession has value
\be
-\frac{\l}{(1+\omega^2\rho^2)\rho^2}(\omega^2\rho^2\uc -\Om\q)\land\q=
\frac{\l}{(1+\omega^2\rho^2)}(\Om - \omega^2\uc)\land\q.
\end{equation}

\section{Discussion} 

The systematic use of absolute velocities instead of relative ones
gives us a nice form of the Lorentz boosts which results in 
extremely brief and simple derivation of

-- \ the discrete Thomas rotation due to successive Lorentz boosts,

-- \ the Fermi-Walker equation,

-- \ the Thomas rotation on a circular world line,

-- \ Thomas rotations in general,

-- \ the Thomas precession with respect to an inertial frame,

\noindent and it allows us a deeper insight into the nature of
Thomas rotations and Thomas precessions. It is an important fact that 
the Thomas rotation is absolute i.e. independent of reference frames 
while the Thomas precession is relative i.e. refers to inertial
frames. It is emphasized again that the same
gyroscope shows different precessions to different inertial frames.

REFERENCES	

Costella J.P., McKellar B.H.J., Rawlinson A.A.and Stephenson G.J. (2001)
{\it Am. J. of Phys.} {\bf 69}, 837

Fisher G.P. (1972) {\it Am.J.Phys.} {\bf 40}, 1772

Herrera L. and di Prisco A. (2002) {\it Found.Phys.Lett.} {\bf
15}, 373

W. L. Kennedy (2002) {\it Eur. J. Phys.} {\bf 23} 235

L\'evay P. (2004) {\it J. Phys. A: Math. Gen.} {\bf 37} 4593

Matolcsi T. (1993) {\it Spacetime without Reference Frames}, Akad\'emiai
Kiad\'o Budapest

Matolcsi T. (1998) {\it Found. Phys.} {\bf 27}, 1685

Matolcsi T. (2001) {\it Stud.Hist.Phil.Mod.Phys.} {\bf 32}, 83

M\o ller M.C. {\it The Theory of Relativity} (1972) 2.ed. Oxford,
Clarendon Press

Muller R.A. (1992) {\it Am. J. Phys.} {\bf 60}, 313

Philpott R.J. (1996) {\it Am. J. Phys.} {\bf 64}, 552

Rebilas K. (2002) {\it Am. J. Phys.} {\bf 70}, 1163

Rhodes J.A. and Semon M.D. (2004) {\it Am. J. Phys.} {\bf 72}, 943

Rindler W. and Perlick V. (1990) {\it Gen.Rel.Grav.} {\bf 22}, 1067 

Thomas L. H. (1927) {\it Phil. Mag.} {\bf 3}, 1 

Ungar A. A. (1989) {\it Found. Phys.} {\bf 19}, 1385

Ungar A. A.
{\it Beyond the Einstein addition law and its gyroscopic Thomas precession. 
The theory of gyrogroups and gyrovector spaces.} (2001) Fundamental Theories of Physics, 117, 
Kluwer Academic Publishers Group

 \end{document}